\documentclass[preprint,12pt]{aastex}

\shortauthors{Henry \& Winn~2007}
\shorttitle{Rotation Period of HD~189733}

\begin{document}

%
\def\ltsima{$\; \buildrel < \over \sim \;$}
\def\lsim{\lower.5ex\hbox{\ltsima}}
\def\gtsima{$\; \buildrel > \over \sim \;$}
\def\gsim{\lower.5ex\hbox{\gtsima}}
\def\lam{\lambda=-1\fdg4 \pm 1\fdg1}
%

\bibliographystyle{apj}

\title{
The Rotation Period of the Planet-Hosting Star HD~189733
}

\author{Gregory W.\ Henry}

\affil{Center of Excellence in Information Systems, Tennessee State
  University, 3500 John A.\ Merritt Blvd., Box 9501, Nashville, TN
  37209, USA}

\author{Joshua N.\ Winn}

\affil{Department of Physics, and Kavli Institute for Astrophysics and
  Space Research, Massachusetts Institute of Technology, Cambridge, MA
  02139, USA}

\begin{abstract}

  We present synoptic optical photometry of HD~189733, the
  chromospherically active parent star of one of the most intensively
  studied exoplanets. We have significantly extended the timespan of
  our previously reported observations and refined the estimate of the
  stellar rotation period by more than an order of magnitude: $P =
  11.953\pm 0.009$~days. We derive a lower limit on the inclination of
  the stellar rotation axis of $56\arcdeg$~(with 95\% confidence),
  corroborating earlier evidence that the stellar spin axis and
  planetary orbital axis are well aligned.

\end{abstract}


\section{Introduction}

Of all the known transiting exoplanets, in many ways the best target
for intensive study is HD~189733b, discovered by Bouchy et
al.~(1995). The star is nearby and bright ($d=19$~pc, $V=7.7$); the
planet-to-star contrast is relatively large ($R_p/R_\star = 0.16$);
and the orbital period is relatively short (2.2~days). Thus, the star
grants numerous opportunities to observe deep transits with a high
signal-to-noise ratio. Investigators have seized these opportunities
to measure the stellar and planetary radii (Bakos et al.~2006, Winn et
al.~2007, Baines et al.~2007, Pont et al.~2007), the alignment between
the stellar rotation and orbital axes (Winn et al.~2006), the planet's
infrared spectrum (Deming et al.~2007, Grillmair et al.~2007), the
planet's infrared brightness distribution (Knutson et al.~2007), and
the planetary atmosphere (Tinetti et al.~2007).

However, one aspect of HD~189733 is not ideal: the star is relatively
active, with strong chromospheric emission (Wright et al.~2004) and a
spotted photosphere (Winn et al.~2007; Pont et al.~2007). This
activity effectively causes extra noise of $\approx$12~m~s$^{-1}$ in
Doppler velocities and $\approx$1\% in relative photometry with a
characteristic timescale equal to the stellar rotation period, which
has been estimated by various means to be 11-13~days (Bouchy et
al.~2005, H\'ebrard \& Lecavelier des \'Etangs 2006, Winn et
al.~2007). Because the rotation period is much longer than the transit
duration (1.8~hr), the activity-related noise can often be neglected
in a single-transit study, or it can be accounted for with a simple
prescription such as fitting a linear function of time to the
out-of-transit Doppler velocities or the relative flux. However, a
more sophisticated understanding of the stellar variability is
important for very high-precision work and for comparing observations
of different transits. For example, the discrepant values for
$R_p/R_\star$ that have been obtained by different authors might be
explained as the result of spot-pattern variability.

Although the photospheric variations can be a nuisance for
high-precision work, they do provide at least one benefit, namely, the
possibility of measuring the stellar rotation period with great
accuracy. If the spot pattern is relatively stable over several
rotation cycles, then a quasiperiodicity may be detectable in the
stellar flux. The rotation period is a fundamental quantity
characterizing the star as well as the angular momentum evolution of
the planetary system, and the period can also be used to study the
alignment between the stellar rotation axis and the orbital axis (Winn
et al.~2007).

In this paper we present optical photometry of HD~189733 spanning
three observing seasons, from 2005 October to 2007 July.  Our goal is
to provide a fuller context for some of the specific datasets that
were taken during this timespan, and to help in planning future
investigations by setting realistic expectations for the typical
variability of this star. In the following section, we describe the
observations and the data and provide the entire data set in
electronic form. In \S~3, we use the data to determine the stellar
rotation period with much greater precision than has previously been
possible. [A portion of the data was presented by Winn et al.~(2007),
but the expanded dataset presented here allows for tremendous
improvement.]  In \S~4, we use the improved stellar rotation period to
update the determination of the inclination angle of the stellar
rotation axis with respect to the sky plane. The last section
summarizes the results.

\section{Observations and Data Reduction}

We acquired 314 good observations of HD~189733 between 2005 October
and 2007 July with the T10 0.8m automated photometric telescope (APT)
at Fairborn Observatory in southern Arizona. The APT uses two
temperature-stabilized EMI 9124QB photomultiplier tubes to detect
photon count rates simultaneously through Str\"omgren $b$ and $y$
filters. On a given night, the telescope automatically makes
observations of each target star and three nearby comparison stars,
along with measures of the dark count rate and sky brightness in the
vicinity of each star.  Designating the comparison stars as A, B, and
C, and the target star as D, the observing sequence is DARK, A, B, C,
D, A, SKY$_{\rm A}$, B, SKY$_{\rm B}$, C, SKY$_{\rm C}$, D, SKY$_{\rm
  D}$, A, B, C, D. In this case, the target star D was HD~189733, and
the comparison stars A, B, and C were HD~189410, HD~191260, and
HD~189410, respectively (i.e., we chose to make duplicate observations
of HD~189410 rather than use a third comparison star). A diaphragm
size of $45\arcsec$ and an integration time of 20 seconds were used
for all integrations.

The observations were reduced to form 3 independent measures of each
of the 6 differential magnitudes D$-$A, D$-$B, D$-$C, C$-$A, C$-$B,
and B$-$A. These were placed on the standard Str\"omgren system with
yearly mean transformation coefficients. These differential magnitudes
were corrected for dead time and differential extinction. To increase
the signal-to-noise ratio, the data from the $b$ and $y$ passbands
were averaged to create ``$(b+y)/2$'' magnitudes. After passing
quality control tests, the 3 independent measures of each differential
magnitude were combined, giving one mean data point per complete
sequence for each of the 6 differential magnitudes. The standard
deviations of a single observation about the means of the 314 D$-$A,
D$-$B, D$-$C, C$-$A, C$-$B, and B$-$A differentials are 0.0064,
0.0064, 0.0064, 0.0014, 0.0019, and 0.0018, respectively. Because C
and A represent the same star, the C$-$A standard deviation of
0.0014~mag is a good estimate of the limiting uncertainty in each
measurement. The C$-$B and B$-$A standard deviations of 0.0019 and
0.0018~mag show that the two comparison stars (A and B) are
intrinsically constant to 0.001~mag or better. The much larger
standard deviations (0.0064~mag) of the D$-$A, D$-$B, and D$-$C data
indicate that HD~189733 is significantly variable. For additional
information on the telescopes, photometers, observing procedures, data
reduction techniques, and photometric precision, see Henry~(1999) or
Eaton, Henry, \& Fekel (2003).

The individual $(b+y)/2$ observations for all 6 combinations of
differential magnitudes are given in Table~1. The time variation of
the D$-$A observations is evident in the top panel of Figure~1. The
time series can be divided into 4 segments of nearly nightly coverage
interrupted by long gaps. The first segment is the latter half of the
2005 observing season. The second and third segments are from the 2006
observing season, which is interrupted by the summer shutdown of the
APTs during Arizona's rainy season. The fourth segment is from the
first half of the 2007 observing season up until the 2007 summer
shutdown.

It is clear from Figure~1 that HD~189733 was variable throughout our
entire photometric campaign.  As noted above, the standard deviation
of the complete D$-$A data set is 0.0064 mag.  The standard deviations
of the 4 contiguous time ranges shown in the lower 4 panels of Fig.~1
are (from top to bottom) 0.0058, 0.0044, 0.0083, and 0.0053 mag.  From
the beginning until approximately JD~2,453,880 (late in May of 2006
and marked with an arrow in Fig.~1), the variability was rather
erratic.  However, from that time forward, the star has exhibited
coherent oscillations until the end of our observations.

\begin{figure}[p]
\epsscale{0.75}
\plotone{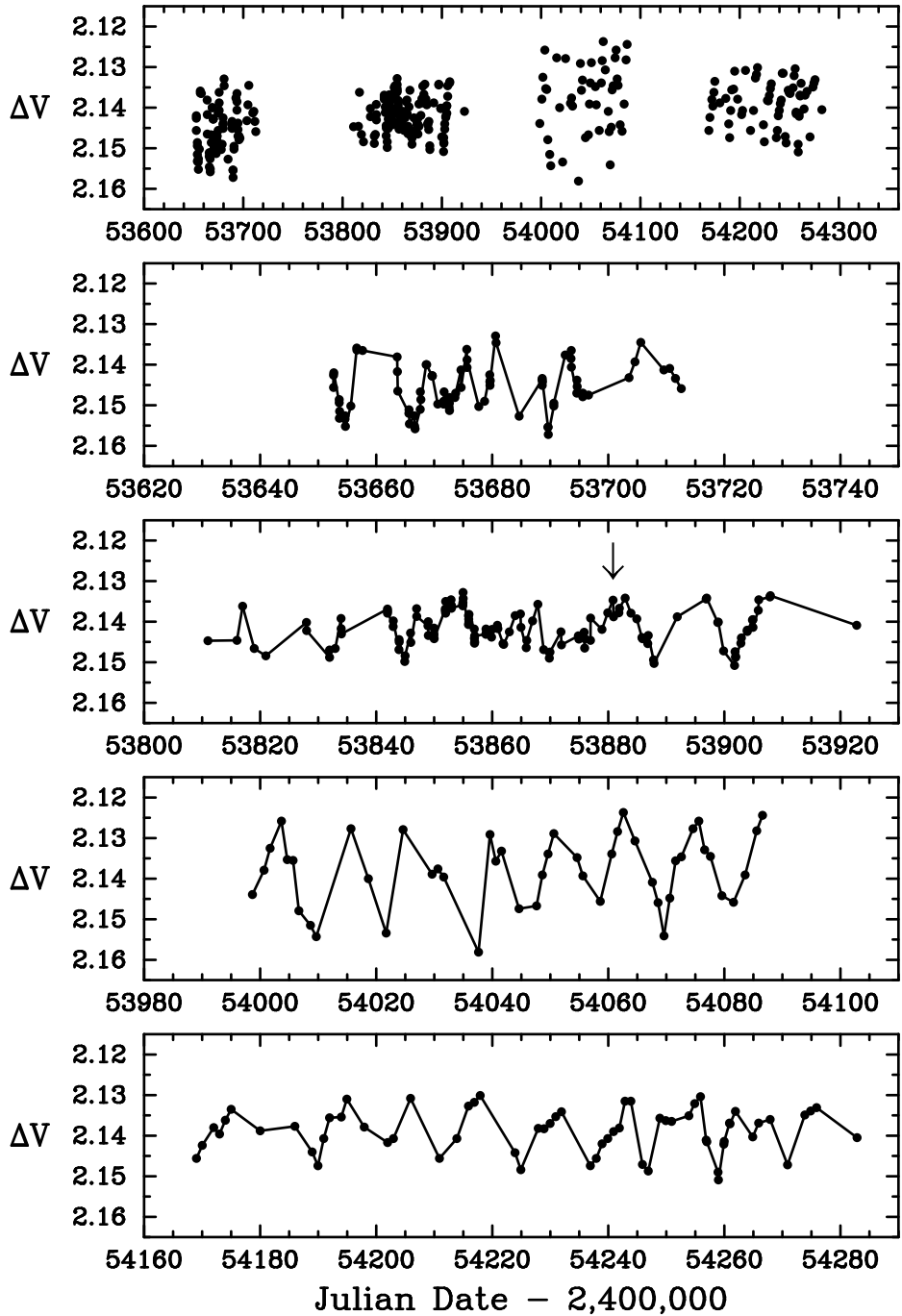}
\caption{
Photometric variability of HD~189733, as observed with
the T10 0.8m APT in the $(b+y)/2$ passband. The magnitude
difference between HD~189733 and the comparison star 
HD~189410 is plotted. The top panel shows the
entire data set, while the other panels focus
on each of the four segments of near-nightly observations.
\label{fig:1}}
\end{figure}

\section{Results}

Chromospherically active stars are often observed to exhibit erratic
and occasionally quasiperiodic photometric variability (see, e.g.,
Vaughan et al.~1981; Henry, Fekel, \& Hall 1995; Lockwood et
al.~2007). The quasiperiodic component of the variability is
interpreted as the result of photospheric spots and plages that are
carried into and out of view as the star rotates (Wilson 1978). The
coherence of the oscillations is limited by time variations of the
spot pattern on the rotational timescale, as well as emission
variations on shorter timescales. Adopting the same interpretation for
HD~189733, the spots cover $\sim$1\%--2\% of the stellar
surface at any moment. Their distribution is apparently complex but
was reasonably stable between 2006 May and 2007 July, the time period
during which coherent oscillations were observed (see Fig.~1).

This allows us to estimate the rotation period of HD~189733 with
greater precision than previous estimates. The top panel of Fig.~2
shows the frequency power spectrum of the D$-$A data from JD~2,453,880
onward, computed with the method of Van{\'{\i}}{\v c}ek~(1971). There
is a strong peak corresponding to a period of 11.952~days with an
estimated uncertainty of 0.016~days. Similar analyses of the D$-$B and
D$-$C data (the differential magnitudes derived from different
comparison observations) gave $11.952 \pm 0.017$~days and $11.955 \pm
0.016$~days, respectively.  These three estimates are all in good
agreement. The mean and standard deviation of the mean of these three
period determinations give $P = 11.9527\pm0.0010$ days, but this
precision is probably too optimistic since the three period
determinations are not completely independent. Therefore, as our best
estimate of the rotation period, we adopt the weighted mean of the
three determinations and its formal uncertainty: $P =
11.953\pm0.009$~days. The phase-folded light curve is shown in the
bottom panel of Fig.~2. The quasi-sinusoidal pattern is typical of
spot-induced variability with modest cycle-to-cycle spot variations
(Henry, Fekel, \& Hall 1995).

How does this compare to previous estimates of the rotation period?
Bouchy et al.~(2005) estimated $P\sim 11$~days (but did not venture an
uncertainty), based on the combination of the estimated stellar
radius, the measurement of the stellar projected rotation velocity
$v\sin i$, and the assumption $\sin i = 1$. H\'ebrard \& Lecavelier
des \'Etangs~(2006) analyzed the {\it Hipparcos}\, photometry for this
star and found candidate periods at 13.3, 11.8, 8.8, and 4.6~days.
They identified the 11.8-day period as the likely rotation period but
did not quantify the uncertainty, as their main concern was
identifying planetary transits and refining the estimate of the {\it
  orbital}\, period. A portion of our data were published previously
(Winn et al.~2007), yielding the estimate $P=13.4\pm 0.4$~days, but
that estimate was based on the observation of only 2.5 cycles over
approximately 40 days.\footnote{In Fig.~2, the portion that was
  previously analyzed ranged from the arrow in the central panel to
  the right-hand edge of the central panel.} This is why we cautioned
that continued monitoring was needed to check on the period
estimate. In contrast, during the full campaign presented in this
paper, we observed 18 complete cycles over approximately 400 days.

\begin{figure}[p]
\epsscale{0.9}
\plotone{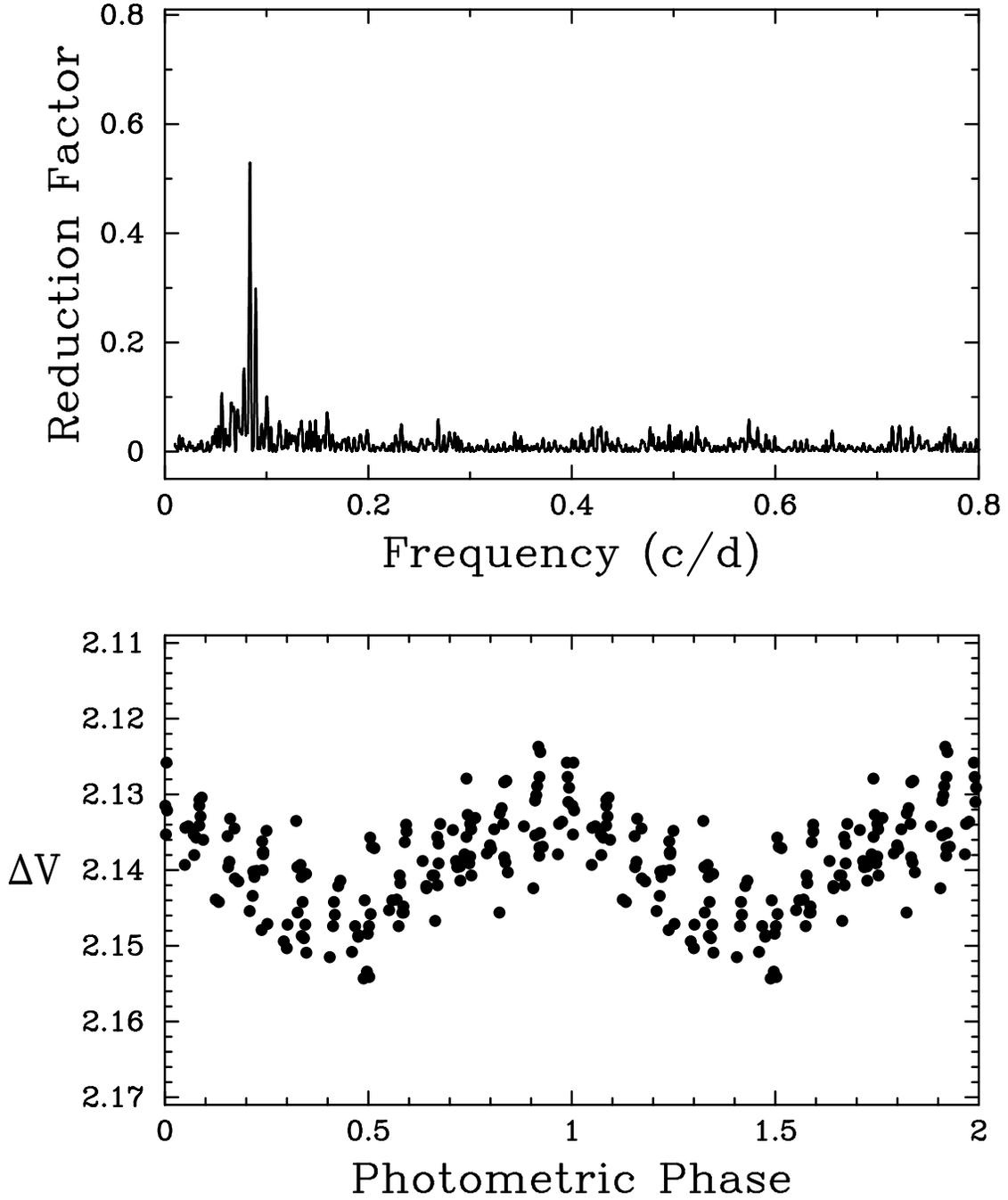}
\caption{
(Top). Frequency power spectrum of the D~$-$~A data,
computed with the method of Van{\'{\i}}{\v c}ek~(1971).
(Bottom). Phase-folded light curve
using the period $P=11.953$~days.
\label{fig:2}}
\end{figure}

\section{Spin-Orbit Alignment}

As alluded to earlier, the measured rotation period can be used to
estimate the inclination of the stellar rotation axis with respect to
the sky plane. Since the planetary orbit is nearly edge-on, it is
natural to suppose the stellar equator is also viewed nearly edge-on,
and it is interesting to test this hypothesis. Even if planet
formation usually results in the alignment of planetary orbital axes
and stellar rotation axis---an assumption that itself is worth
checking---some of the planetary ``migration'' mechanisms that have
been proposed to explain close-in giant planets such as HD~189733
could produce large misalignments. In particular, a broad distribution
of spin-orbit angles is predicted by theories involving planet-planet
scattering (Chatterjee et al.~2007) and Kozai cycles with tidal
friction (Fabrycky \& Tremaine 2007).

There are at least two observable aspects of spin-orbit alignment. The
first aspect is the Rossiter-McLaughlin effect, a spectroscopic
distortion that occurs during transits. Observations of this
``spectroscopic transit'' can be used to estimate the angle $\lambda$
between the {\it sky projections}\, of the stellar rotation axis and
the orbital axis (Ohta et al.~2005, Gaudi \& Winn 2007). For this
system, $\lambda$ has been measured to be $-1.4\arcdeg \pm 1.1\arcdeg$
(Winn et al.~2006).\footnote{Incidentally, the observation that the
  rotation period is much longer than the orbital period of 2.2~days
  implies that the two periods have not been tidally
  synchronized. Since the expected timescale for reorientation of the
  star is of the same order of magnitude as the timescale for
  synchronization, this finding supports the interpretation of the
  observed spin-orbit alignment as primordial, rather than the outcome
  of tidal effects. For further discussion, see Fabrycky et
  al.~(2007).} The second aspect is the inclination $i$ of the stellar
rotation axis with respect to the sky plane, which can be written in
terms of measurable quantities as follows:
\begin{equation}
\sin i = v\sin i \left( \frac{P}{2\pi R_\star} \right).
\end{equation}
This calculation was done by Winn et al.~(2007) using the data
available at that time. We update the result using the new value $P=
11.953\pm 0.009$~days as well as the more precise value $R_\star =
0.755\pm 0.011~R_\odot$ that has become available through space-based
photometry (Pont et al.~2007). The most precise measurement of $v\sin
i$ remains unchanged, $2.97\pm 0.22$~km~s$^{-1}$ (Winn et al.~2006).
Assuming Gaussian errors in all these quantities, the result is $\sin
i = 0.93\pm 0.07$.

We estimated the {\it a posteriori}\, probability distribution $p(i)$
as follows. For each trial value of $i$ from $0\arcdeg$ to
$90\arcdeg$, we computed
\begin{equation}
\chi^2 = \left( \frac{\sin i - 0.93}{0.07} \right)^2
\end{equation}
and made the usual maximum-likelihood assumption $p(i) \propto
\exp(-\chi^2/2)$. The resulting probability distribution and the
corresponding cumulative distribution are shown in Fig.~3.  Although
the most probable value of $i$ is $68\arcdeg$, more edge-on
configurations are not strongly disfavored, and hence the results are
best interpreted as a lower limit on the inclination angle: $i >
56\arcdeg$ with 95\% confidence. This lower bound is hardly changed at
all from the previous result of Winn et al.~(2007), and is indeed
slightly {\it less}\, stringent. The limiting factor is the
uncertainty in $v\sin i$, and it is not obvious how further
improvement can be made, since the uncertainty is dominated by
calibration error rather than random error.

\begin{figure}[p]
\epsscale{0.5}
\plotone{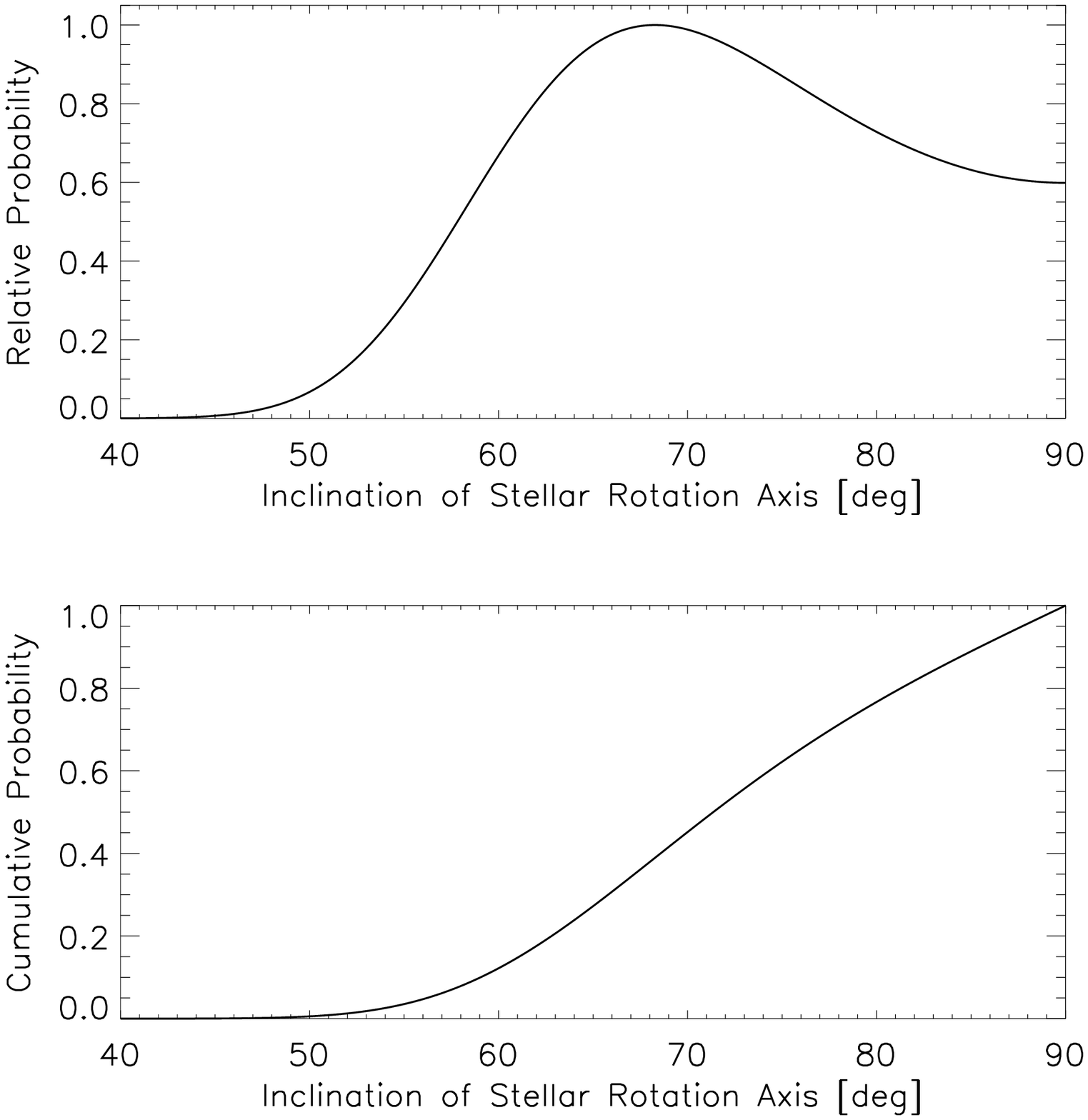}
\caption{ (Top). The {\it a posteriori}\, probability distribution for
  $i$, the inclination of the stellar rotation axis with respect to
  the sky plane.  (Bottom). The corresponding cumulative probability
  distribution.  The 95\% lower limit on $i$ is $59\arcdeg$.
  \label{fig:3}}
\end{figure}

\section{Summary}

We have presented optical photometry spanning 3 observing seasons for
HD~189733, one of the very brightest stars that has a known transiting
planet. During the latter half of our campaign, we detected periodic
photometric variations that allowed us to refine the estimated stellar
rotation period by more than an order of magnitude. Our photometric
data will allow the numerous investigators who carried out intensive
transit observations between 2005 and 2007 to assess any spot-induced
effects on their data and will help to set realistic expectations for
future programs. We confirm the previously published result that the
stellar rotation axis is not strongly misaligned with the orbital
axis. Given that the intense interest in this object does not seem to
be subsiding, we plan to continue the APT observations over the next
few seasons.

\acknowledgments We thank E.~Turner, B.~Croll, and J.~Mathews for
helpful discussions.  GWH acknowledges support from NSF grant
HRD-9706268 and NASA grant NNX06AC14G.

\begin{deluxetable}{lrrrrrr}
\tabletypesize{\normalsize}
\tablecaption{Individual Photometric Observations of HD~189733\label{tbl:photometry}}
\tablewidth{0pt}

\tablehead{
\colhead{Hel. Julian Date} & \colhead{D$-$A} & \colhead{D$-$B} & \colhead{D$-$C} & \colhead{C$-$A} & \colhead {C$-$B} & \colhead {B$-$A} \\
\colhead{(HJD $-$ 2,400,000)} & \colhead{(mag)} & \colhead{(mag)} & \colhead{(mag)} & \colhead{(mag)} & \colhead{(mag)} & \colhead{(mag)}
}

\startdata
53652.6447 & 2.1426 & 0.4843 & 2.1406 &    0.0020 & $-$1.6563 & 1.6583 \\
53652.6652 & 2.1456 & 0.4858 & 2.1469 & $-$0.0014 & $-$1.6611 & 1.6597 \\
53652.6932 & 2.1420 & 0.4833 & 2.1425 & $-$0.0005 & $-$1.6592 & 1.6588 \\
53653.6206 & 2.1494 & 0.4920 & 2.1505 & $-$0.0012 & $-$1.6586 & 1.6574 \\
53653.6416 & 2.1486 & 0.4915 & 2.1514 & $-$0.0027 & $-$1.6598 & 1.6570 \\
\enddata

\tablecomments{Differential magnitudes were measured in the
  Str\"omgren $(b+y)/2$ passband.  Table~1 is presented in its
  entirety in the electronic edition of this journal.  A portion is
  shown here for guidance regarding its form and content.}

\end{deluxetable}

\end{document}